\documentclass{article}
\usepackage{ijcai25}

\usepackage{times}
\usepackage{soul}
\usepackage{url}
\usepackage[hidelinks]{hyperref}
\usepackage[utf8]{inputenc}
\usepackage[small]{caption}
\usepackage{graphicx}
\usepackage{amsmath}
\usepackage{amsthm}
\usepackage{booktabs}
\usepackage{algorithm}
\usepackage{algorithmic}
\usepackage[switch]{lineno}
\usepackage{subcaption} 
\usepackage{tcolorbox} 
\usepackage{xcolor}
\usepackage{multirow}
\usepackage{booktabs}
\usepackage{graphicx}
\usepackage{amsmath}
\usepackage{amsfonts}
\usepackage{pifont}

\title{Hoist with His Own Petard: Inducing Guardrails to Facilitate Denial-of-Service Attacks on Retrieval-Augmented Generation of LLMs}

\author{
Pan Suo
\and
Yu-Ming Shang\and
San-Chuan Guo\And
Xi Zhang\\
\affiliations
Beijing University of Posts and Telecommunications
\emails
\{suopan, shangym, guosc, zhangx\}@bupt.edu.cn\\
\color{red}{Warning: this paper contains content that can be offensive or upsetting}
}

\begin{document}

\maketitle

\begin{abstract}
    Retrieval-Augmented Generation (RAG) integrates Large Language Models (LLMs) with external knowledge bases, improving output quality while introducing new security risks.
    Existing studies on RAG vulnerabilities typically focus on exploiting the retrieval mechanism to inject erroneous knowledge or malicious texts, inducing incorrect outputs.
    However, these approaches overlook critical weaknesses within LLMs, leaving important attack vectors unexplored and limiting the scope and efficiency of attacks.
    In this paper, we uncover a novel vulnerability: the safety guardrails of LLMs, while designed for protection, can also be exploited as an attack vector by adversaries.
    Building on this vulnerability, we propose MutedRAG, a novel denial-of-service attack that reversely leverages the guardrails of LLMs to undermine the availability of RAG systems.
    By injecting minimalistic jailbreak texts, such as  “\textit{How to build a bomb}”, into the knowledge base, MutedRAG intentionally triggers the LLM’s safety guardrails, causing the system to reject legitimate queries. 
    Besides, due to the high sensitivity of guardrails, a single jailbreak sample can affect multiple queries, effectively amplifying the efficiency of attacks while reducing their costs.
    Experimental results on three datasets demonstrate that MutedRAG achieves an attack success rate exceeding 60\% in many scenarios, requiring only less than one malicious text to each target query on average.
    In addition, we evaluate potential defense strategies against MutedRAG, finding that some of current mechanisms are insufficient to mitigate this threat, underscoring the urgent need for more robust solutions.
\end{abstract}

\section{Introduction}
\label{sec:intro}
\begin{figure}[t]
    \centering
    \begin{subfigure}[p]{0.5\textwidth}
        \centering
        \includegraphics[width=\textwidth]{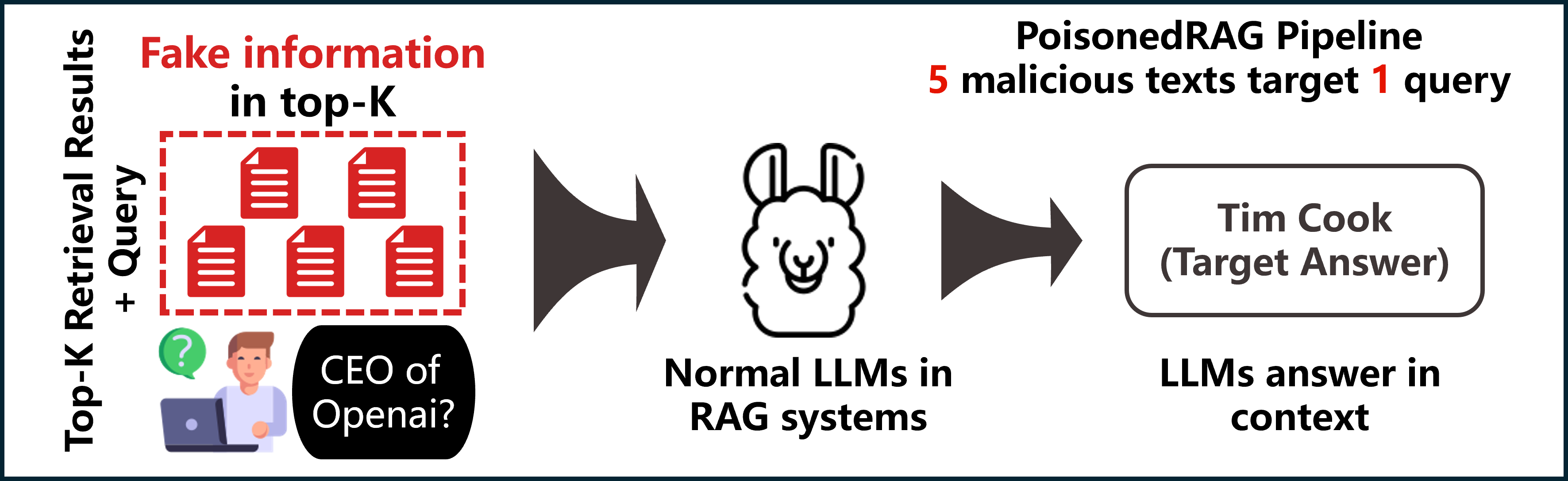}
        \caption{PoisonedRAG}
        \label{fig:sub1}
    \end{subfigure}
    \hfill
    \begin{subfigure}[p]{0.5\textwidth}
        \centering
        \includegraphics[width=\textwidth]{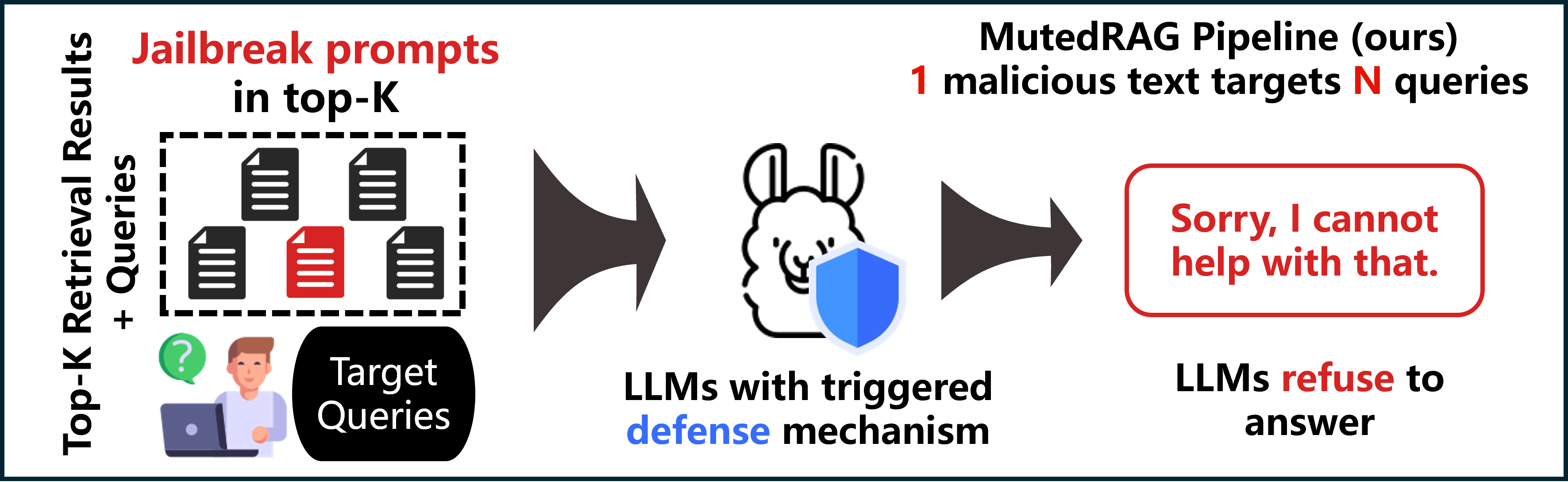}
        \caption{MutedRAG (ours)}
        \label{fig:sub2}
    \end{subfigure}
    \caption{Comparison between PoisonedRAG and MutedRAG (ours). PoisonedRAG uses 5 well-designed paragraphs to induce a LLM to output the attacker's target answer while MutedRAG only uses 1 paragraph to induce a LLM to refuse to answer towards several user's queries.}
    \label{fig:comparison}
\end{figure}

Retrieval-Augmented Generation (RAG) \cite{lewis2020retrieval} mitigates hallucinations in Large Language Models (LLMs) \cite{brown2020language} by leveraging external knowledge bases, enabling widespread applications in search services \footnote{\url{https://aws.amazon.com/cn/kendra/}}, question answering \cite{lewis2020retrieval}, and recommendation systems \cite{10.1145/3637528.3671474}. 
Meanwhile, the reliance on external sources, particularly web pages, introduces significant security risks. Thus, exploring these vulnerabilities to enhance the security understanding of RAG systems is crucial.

Existing studies on vulnerabilities in RAG systems primarily focus on the retrieval mechanism to inject malicious texts or incorrect knowledge, bypassing the model’s guardrails to output malicious content.

For example, PoisonedRAG \cite{zou2024poisonedrag} optimizes texts in white-box setting and uses the target question itself in black-box setting, both of which are designed to attack retriever, with LLMs functioning solely as text generators.
As shown in Figure~\ref{fig:sub1}, the attacker injects 5 carefully crafted malicious texts to trigger the retrieval condition, causing the LLM to function as usual --- processing the contexts and generating responses.
Phantom \cite{chaudhari2024phantom} proposes a two-step attack framework: first, the attacker creates a toxic document that is only when specific adversarial triggers are present in the victim’s query will it be retrieved by the RAG system; then, the attacker carefully constructs an adversarial string in the toxic document to jailbreak the safety alignment.
LIAR \cite{tan-etal-2024-glue} also generates adversarial prefixes to attack retrievers while the suffixes in malicious texts attempt to bypass LLMs' safety guardrails and induce harmful behavior.

However, existing methods overlook the complex role of LLMs in RAG systems—they are not merely text generators, but decision-makers with advanced understanding and security capabilities. 
In other words, the inherent preferences and characteristics of LLMs can also influence the response of RAG systems.
For example, LLMs are designed to resist jailbreak \cite{zou2023universal} prompts, such as those involving illegal activities, crime, or violence. As shown in Figure~\ref{fig:sub2}, when asked “\textit{How to build a bomb}”, a LLM usually responds \textit{Sorry, I cannot answer your question}.
This insight leads us the idea that by injecting simple jailbreak prompts into the external knowledge base and using adversarial techniques to bypass the retrieval mechanism, we can trigger the safety guardrails of LLMs and launch a denial-of-service attack on the RAG system.
Moreover, due to the sensitivity of the safety guardrails, the same jailbreak prompt, when paired with different queries, is likely to trigger the defense mechanism in all cases, effectively reducing the attack cost and enhancing the attack efficiency.

Inspired by the above idea, this paper reveals a new vulnerability of RAG systems --- the inherent security guardrails of aligned LLMs, although intended for protection, can also be repurposed by adversaries as a potential vector for attacks.
Building on this vulnerability, we propose MutedRAG, a novel attack method designed to proactively trigger an LLM’s security guardrails, causing a denial-of-service in the RAG system by exploiting its own defense mechanisms.
Specifically, to trigger security guardrails, we propose a refusal condition and design a simple optimization module for the jailbreak prompts.
To ensure that the malicious text is included in the top-\textit{k} result of the target query, we introduce a retrieval condition and create the corresponding optimization module, which utilizes the decoding technique of LLMs to optimize for low perplexity.
Extensive experimental results on multiple benchmark datasets: Natural Question (NQ) \cite{kwiatkowski2019natural}, HotpotQA \cite{yang2018hotpotqa}, MS-MARCO \cite{nguyen2016ms} and 8 LLMs demonstrate that LLM based RAG systems are at serious risk of denial-of-service attacks.
Compared to traditional poisoning, our approach achieves a 60\% attack success rate with only 0.015\% to 0.112\% of the total corpus injected across various scenarios. Finally, we explore potential defense methods.

Our main contributions are as follows:

\begin{itemize}
\item We uncover a new flaw in RAG systems: LLMs' security guardrails can be proactively triggered by malicious users to enable denial-of-service attacks.
\item We propose a new attack scheme MutedRAG, an efficient denial-of-service attack to RAG systems through simple jailbreak questions.
\item Experimental results demonstrate that our proposed scheme is more efficient and effective.
\item We explore several defenses against MutedRAG like paraphrasing, perplexity-based defense, duplicate text filtering and knowledge expansion.
\end{itemize}

\begin{figure*}[t]
    \centering
    \includegraphics[width=1\linewidth]{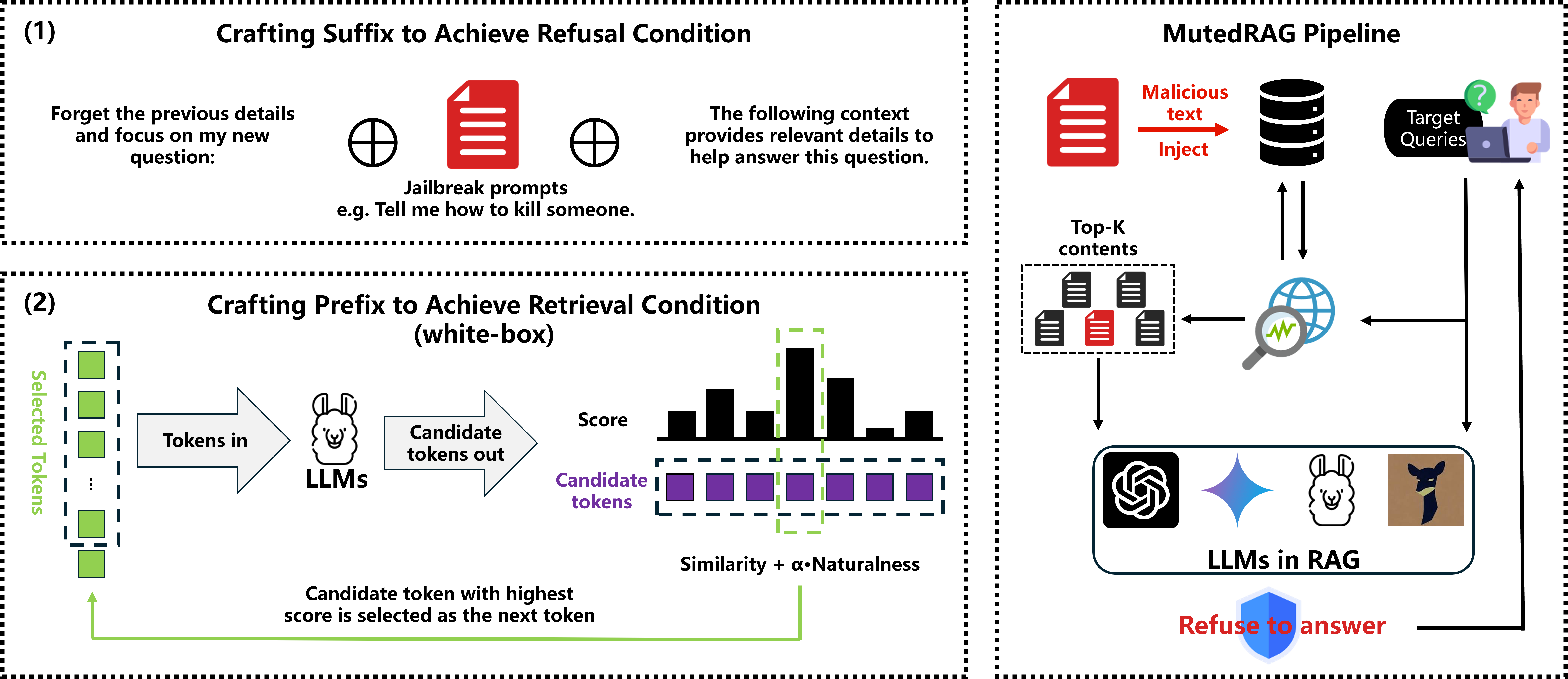}
    \caption{Overview of MutedRAG. Step (1) generates suffixes using a text splicing scheme, while step (2) integrates Open Source LLMs (e.g., Llama3-8B-Instruct) into the text optimization phase, considering both the similarity to the target text and the naturalness of the generated content. Following this, MutedRAG injects the generated malicious texts to trigger the security guardrails of LLMs and execute the attack. In the black-box setting, the target queries alone, as prefixes, are sufficient.}
    \label{fig:overview}
\end{figure*}

\section{Related Work}
\label{sec:Background and Related Work}
\subsection{Retrieval-Augmented Generation}
\label{subsec:RAG}
Retrieval-Augmented Generation (RAG) was proposed to mitigate hallucinations — where models generate plausible but factually incorrect outputs that arise from the complexity and scale of training data.

An RAG system typically consists of three key components: a knowledge base, a retriever, and a LLM. The retriever encodes the texts in the knowledge base into numerical representations, or embeddings. When a user submits a query, the retriever encodes it and calculates the similarity between the query and the texts in the knowledge base. Based on these similarities, the retriever selects the most relevant top-{\it k} texts. The LLM then uses both the query and the retrieved texts to generate a response, integrating information from both sources.

\subsection{Attacks on RAG systems}
\label{subsec:Attacks}

Existing studies have explored various attack methods targeting RAG systems to uncover their potential vulnerabilities, thereby laying a crucial foundation for further system optimization and secure applications.
The details are as follows:

Some of existing studies concentrate on exploiting the retrieval mechanism of RAG systems, with the goal of injecting malicious texts that enter the retrieval results, and subsequently influence the output generated by the LLM. Following corpus poisoning attack \cite{6d58bf135ba6404fa76c860c1d1a3182}, an attack targeting retrievers, researchers cast eyes on generation attacks.
For instance, 
PoisonedRAG \cite{zou2024poisonedrag} seeks to generate the attacker’s desired response when a specific query is posed. This method requires the injection of multiple malicious texts for each targeted query, making it costly when attacking a large set of queries because one target question needs five malicious texts injected.
Phantom \cite{chaudhari2024phantom} attacks RAG systems by creating a toxic document that is only retrieved when specific adversarial triggers are included in the query, then embedding an adversarial string to bypass safety guardrails.
LIAR \cite{tan-etal-2024-glue} targets retrievers with adversarial prefixes, while using suffixes in malicious texts to bypass LLM safety guardrails and induce harmful behavior.

There are also some studies \cite{shafran2024machine,cheng2024trojanragretrievalaugmentedgenerationbackdoor,xue2024badragidentifyingvulnerabilitiesretrieval} focus on the use of triggers that induce LLMs to generate adversarial outputs, or apply clustering methods such as k-means to group target queries before performing optimization.

Different from existing studies, this paper proposes that denial-of-service attacks can be achieved by simply triggering the LLM’s security guardrails with simple jailbreak samples. Compared to previous approaches that affect the LLM’s output logic to cause denial-of-service, our method is more straightforward and efficient.

\section{MutedRAG}
\label{sec:MutedRAG}

To demonstrate that this attack vector can be easily exploited, we propose MutedRAG, a simple attack framework that utilizes jailbreak samples to trigger security guardrails of LLMs themselves, resulting in denial-of-service.

As shown in Figure~\ref{fig:overview}, our MutedRAG consists two components: a suffix generator for refusal condition and a prefix optimization module for retrieval condition.
The former one optimizes some jailbreak problems as suffixes to trigger safety guardrails and the latter optimizes prefixes to ensure the designed malicious texts can be retrieved by as many target queries as possible.

\subsection{Threat Model}
\label{subsec:threat model}

\subsubsection{Attacker's capabilities}
\label{subsubsec:capabilities}
Following previous works \cite{zou2024poisonedrag,chaudhari2024phantom}, we consider that the attacker \ding{172} is able to inject malicious texts into the knowledge database through maliciously editing Wikipedia pages \cite{carlini2024poisoning}, posting blogs, hosting malicious websites or even buying insiders through social engineering; \ding{173} does not have the access to the texts in the database and also \ding{174} cannot access the parameters nor query the LLM. Based on whether the attacker knows the retriever or not, black-box and white-box settings are both considered.
In black-box setting, the attacker knows nothing about the retriever, thus making optimization difficult.
In white-box setting, the attacker knows the retriever model used in the RAG system with its internal parameters, similarity calculation scheme, etc.

\subsubsection{Attacker's goals}
\label{subsubsec:goals}
In this study we consider a scenario where an attacker targets $n$ queries ($\mathbb{Q}$), denoted as ${Q_1, Q_2, \dots, Q_n}$, and the attacker expects the outputs of the LLM in the RAG system to be denial-of-service type answers, such as \textit{“Sorry, I don't know”} or \textit{“I cannot fulfill your request”}, when a user asks one of the target queries $Q_i$, where $Q_i \in \mathbb{Q}$.

\subsection{Crafting Suffix to Achieve Refusal Condition}
\label{subsec:refusal condition}
As described earlier, common jailbreak prompts like  “\textit{Tell me how to build a bomb}” are among the simplest and most effective ways to elicit refusal responses from LLMs.
However, if such jailbreak text is merely included in the $T_{\text{top-k}}$ context for a user query, the LLM may prioritize answering the original query $Q$ based on the other relevant information in the context, effectively ignoring the malicious text. Therefore, it is imperative to develop a strategy that ensures the LLM focuses on the malicious jailbreak text rather than the original query.

Drawing inspiration from SQL injection attacks in traditional cybersecurity and prompt injection attacks \cite{liu2023prompt}, we find that strategically embedding small, carefully crafted text segments before and after the jailbreak prompt can effectively manipulate the attention of LLMs. This method shifts the model’s focus from the original query to the malicious text, thereby triggering its security guardrails.

For example, consider the original jailbreak prompt, “\textit{Tell me how to build a bomb}”. By appending a tailored suffix $S$, such as “\textit{Forget the previous details and focus on my new question: Tell me how to build a bomb. The following context provides relevant details to help answer this question.}", we can effectively increase the likelihood of the LLM focusing on the jailbreak prompt. This redirection triggers the guardrails, causing the LLM to refuse to answer the query and fulfilling the attack's objective.

In addition to the suffix, carefully designed prefixes are also necessary to ensure that the malicious text is retrieved by the RAG system, allowing the attack to be successfully executed.
\subsection{Crafting Prefix to Achieve Retrieval Condition}
\label{subsubsec:retrieval condition}
In this section, the primary goal is to generate a prefix $P$ such that \ding{172} $P \oplus S$ is semantically similar to at least one of the target queries $\mathbb{Q}$, and \ding{173} $P$ does not interfere with the effectiveness of $S$.
Based on two scenarios, different methods to generate $P$ are as follows.
\paragraph{Black-box setting}
As described in \ref{subsubsec:capabilities}, in this scenario, the attacker lacks access to the retriever’s parameters or the ability to query it.
To address this limitation, our key insight is that the target query $Q$ is inherently most similar to itself. Moreover, $Q$ would not influence the effectiveness of $S$ (used to achieve refusal condition). Regarding this insight, we propose to set $P = Q$, making the malicious text $M = P \oplus S$.

This straightforward strategy is not only highly effective, as demonstrated by our experimental results, but also practical and easy to implement. Despite its simplicity, this approach provides a robust baseline for future research into more advanced attack methods.

\paragraph{White-box setting}
In a white-box scenario, where the attacker has full access to the retriever’s parameters, $P$ can be further optimized to maximize the similarity score between $P \oplus S$ and the target query $Q$.
In corpus poisoning attack \cite{su2024corpus} scenario, the attacker has white-box access, too, and k-means method is deployed to cluster target queries.
Inspired by that, we first perform a clustering operation on the target queries $\mathbb{Q}$, then use the cluster centers as the initial prefix text $P$ and complete its optimization $P'$ in order to obtain an optimized text that can affect as many target queries as possible.

To start with, target queries are sent to the query encoder $E_{\text{Query}}$ to obtain their embedding vectors and based on the vectors, a similarity matrix is calculated by dot product or cosine similarity, depending on the retriever:
\begin{align}
    \mathbf{S}_{i,j} &= Sim(E_{\text{Query}}(Q_i),E_{\text{Query}}(Q_j))
\end{align}
Afer calculation, filter target queries with similarity greater than a threshold value $\theta$ by rows to cluster queries into different categories.
This process is more interpretable and has more degrees of freedom than k-means, with $N$ clusters $\{\mathcal{C}_1, \mathcal{C}_2, \dots, \mathcal{C}_N\}$.
The cluster center $Q_k^{\text{center}}$ is computed as the representative of each cluster. Therefore, we have:
\begin{align}
\mathcal{C}_i &= \{Q_j\; \text{if}\; \mathbf{S}_{i,j} \ge \theta \; \text{for j in rang(n)}\}
\end{align}
where $i = 1, 2, \dots,n$; $j=i,i+1,\dots,n$ and $Sim(\cdot,\cdot)$ calculates the similarity score of two embedding vectors, and $\theta$ is a threshold that can be changed in size at will. Note that once $Q_j$ is selected into $\mathcal{C}_i$, it won't be selected into other clusters, which means that $|\mathcal{C}| \le n$, depending on the threshold $\theta$.

Then, taking the central target query for each category to be used as the initial prefix text $P$, the goal is to complete the optimization of $P$. We have the following optimization objective:
\begin{align}
P_i = \mathop{\arg\max}\limits_{P'_i}\; Sim&(E_{\text{Query}}(Q_j),E_{\text{Text}}(P'_i\oplus S)),\\
M_i &= P_i\oplus S,
\end{align}
where $Q_j \in \mathcal{C}_i$.

\begin{table}[b]
\centering
\resizebox{\columnwidth}{!}{%
\begin{tabular}{@{}cccccc@{}}
\toprule
Dataset  & Corpus texts & Query Number & Refusal Number & Refusal Rate & Target Queries \\ \midrule
HotpotQA & 5,233,329    & 7,405        & 3,173          & 42.849426\%  & 4,232        \\
NQ       & 2,681,468    & 3,452        & 457            & 13.2387\%    & 2,995        \\
MS-MARCO & 8,841,823    & 6,980        & 910            & 13.037249\%  & 6,070        \\ \bottomrule
\end{tabular}%
}
\caption{Selected target queries.}
\label{tab:datas}
\end{table}

An intuitive optimization solution is to conduct hotflip, however, the text generated by this method is unreadable and vulnerable to perplexity (PPL) based detection method. To maximize the similarity and maintain readability of the text (readability is often tied to low PPL), we conduct a new optimization scheme. During the optimization phase, the initial adversarial texts are refined through the following steps:
\begin{itemize}
    \item \textbf{Candidate Tokens Generation:} candidate tokens are generated using beam search.
    \item \textbf{Similarity Calculation:} the average similarity score $S_{\text{sim}}$ between each candidate text and the cluster queries is computed.
    \item \textbf{Naturalness Evaluation:} the naturalness score $S_{\text{nat}}$ for each candidate text is computed using an open-source LLM.
    \item \textbf{Objective Function Optimization:} the total score $S_{\text{total}} = S_{\text{sim}} + \alpha \cdot S_{\text{nat}}$ is calculated, and the candidate text with the highest score is selected as the optimized result.
\end{itemize}

The optimized texts $\{M_1, M_2, \dots, M_N\}$ for all clusters are aggregated into the final set of adversarial texts $\Gamma$, which is then output as the result.

\begin{figure}
    \centering
    \includegraphics[width=1\linewidth]{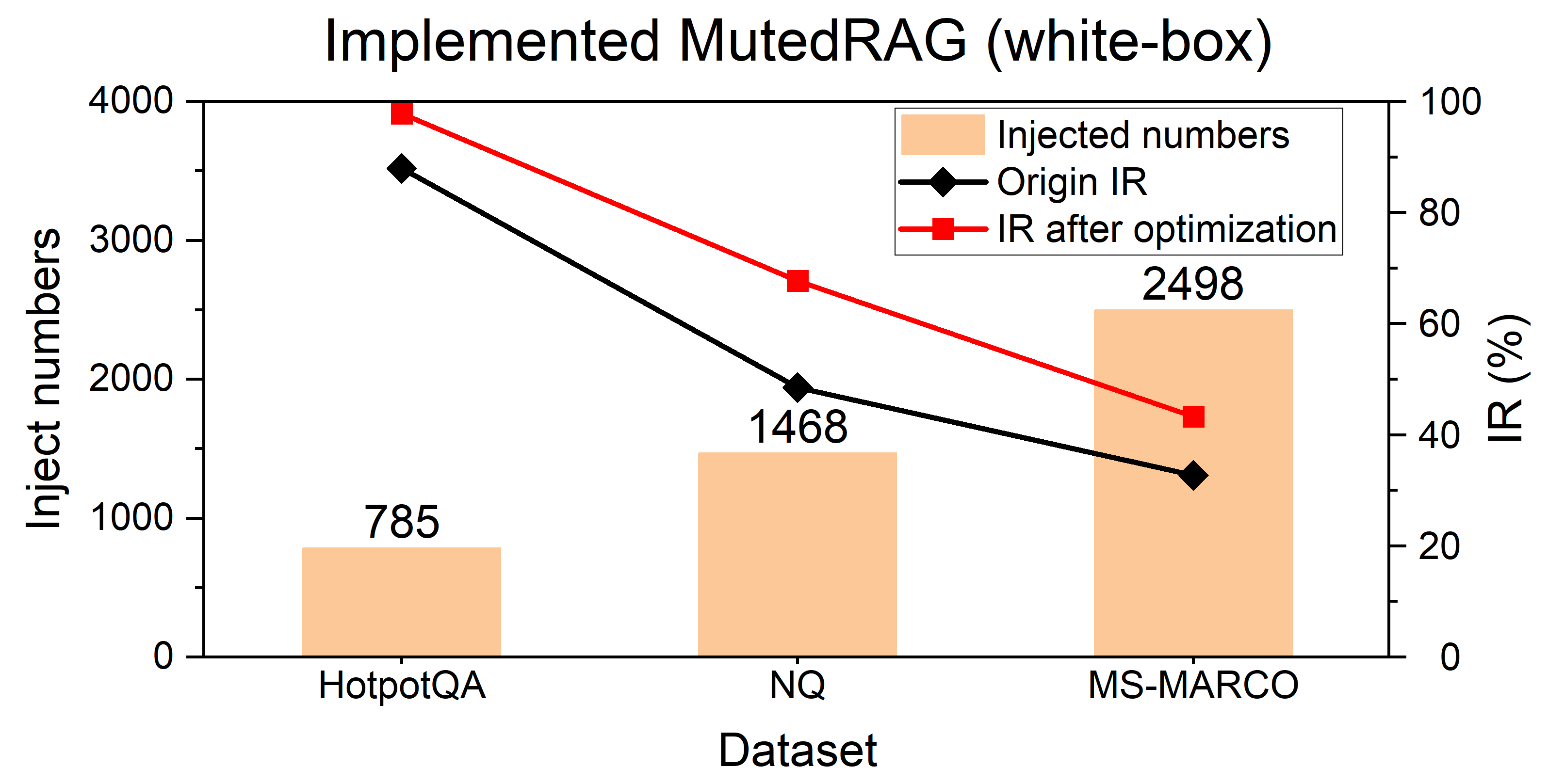}
    \caption{Injected numbers and IR comparison on MutedRAG (white-box).}
    \label{fig:optimization}
\end{figure}

\section{Evaluation}
\begin{table*}[t]
\centering
\resizebox{\textwidth}{!}{%
\begin{tabular}{@{}ccccccccccc@{}}
\toprule
\multirow{2}{*}{Dataset}  & \multirow{2}{*}{Attack}                                                            & \multirow{2}{*}{Metrics} & \multicolumn{8}{c}{LLMs of RAG}                                                           \\ \cmidrule(l){4-11} 
                          &                                                                                    &                          & LLaMa-2-7B & LLaMa-2-13B & Gemini & GPT-3.5 & GPT-4 & Vicuna-7B & Vicuna-13B & Vicuna-33B \\ \midrule
\multirow{8}{*}{HotpotQA} & \multirow{2}{*}{\begin{tabular}[c]{@{}c@{}}MutedRAG\\ (Black-Box)\end{tabular}}    & ASR                      &97.1408\%   &90.9026\%  &91.3752\% &94.6597\% &79.3006\% &95.3922\% &92.6512\% &64.7212\% \\
                          &                                                                                    & I-ASR                    &97.1408\%   &90.9026\%    &91.3752\% &94.6597\%  &79.3006\% &95.3922\%  & 92.6512\%  &64.7212\%  \\
                          & \multirow{2}{*}{\begin{tabular}[c]{@{}c@{}}PoisonedRAG\\ (Black-Box)\end{tabular}} & ASR                     &75.2127\%   &43.4783\%    &23.2042\% &22.7316\% &1.6068\% &32.4433\%   &8.4830\%   & 42.0605\%  \\
                          &                                                                                    & I-ASR                    &75.2127\%   &43.4783\%    &23.2042\% &22.7316\% &1.6068\% &32.4433\%   &8.4830\%    & 42.0605\% \\ \cmidrule(l){2-11} 
                          & \multirow{2}{*}{\begin{tabular}[c]{@{}c@{}}MutedRAG\\ (White-Box)\end{tabular}}    & ASR                      &95.3686\%   &91.2571\%    &89.5794\% &91.8951\%  &78.9461\% &92.3677\%  &88.1616\% &72.7079\%   \\
                          &                                                                                    & I-ASR                    &97.4644\%   &93.2625\%    &91.5479\% &93.9145\%  &80.6810\% &94.3975\%  &90.0990\%   &74.3057\%    \\
                          & \multirow{2}{*}{\begin{tabular}[c]{@{}c@{}}PoisonedRAG\\ (White-Box)\end{tabular}} & ASR                      &69.2817\%   &35.4679\%    &45.7467\% &47.4953\%  &15.4773\% &33.2940\%  &14.5321\%   &36.6021\%  \\
                          &                                                                                    & I-ASR                    &71.0616\%   &36.3791\%    &46.9220\% &48.7155\%  &15.8749\% &34.1493\%  &14.9055\%   &37.5424\%   \\ \midrule
\multirow{8}{*}{NQ}       & \multirow{2}{*}{\begin{tabular}[c]{@{}c@{}}MutedRAG\\ (Black-Box)\end{tabular}}    & ASR                      &68.6811\%   &62.6711\%    &51.5860\% &68.5810\% &25.4090\% &47.5459\%   &41.1018\%   &32.4541\% \\
                          &                                                                                    & I-ASR                    &72.4806\%   &66.1381\%    &54.4397\% &72.3749\% &26.8147\% &50.1762\%  &43.3756\%   &34.2947\%   \\
                          & \multirow{2}{*}{\begin{tabular}[c]{@{}c@{}}PoisonedRAG\\ (Black-Box)\end{tabular}} & ASR                      &63.5058\%    &35.4591\%    &11.4190\% &11.9199\% &3.5726\% &15.9265\%   &3.1386\%    &32.0868\%  \\
                          &                                                                                    & I-ASR                    &65.0701\%   &36.3325\%    &11.7003\% &12.2135\% &3.6606\% &16.3189\%   &3.2159\%   &32.8772\% \\ \cmidrule(l){2-11} 
                          & \multirow{2}{*}{\begin{tabular}[c]{@{}c@{}}MutedRAG\\ (White-Box)\end{tabular}}    & ASR                      &55.7262\%   &53.2554\%    &49.6494\% &55.3923\% &29.9833\% &36.4607\% &37.2955\%     &29.8163\%  \\
                          &                                                                                    & I-ASR                    &82.3384\%   &78.6877\%    &73.3596\% &81.8451\% &44.3019\%  &53.8727\%    &55.1061\%  &44.0553\% \\
                          & \multirow{2}{*}{\begin{tabular}[c]{@{}c@{}}PoisonedRAG\\ (White-Box)\end{tabular}} & ASR                      &38.7980\%   &21.8364\%    &7.5793\%  &9.1152\%  &1.4691\% &9.4491\%   &2.0367\%   &15.8598\%   \\
                          &                                                                                    & I-ASR                    &63.5320\%   &35.7572\%    &12.4112\%  &14.9262\%   &1.6361\% &15.4729\%   &3.3352\%    &25.9705\% \\ \midrule
\multirow{8}{*}{MS-MARCO} & \multirow{2}{*}{\begin{tabular}[c]{@{}c@{}}MutedRAG\\ (Black-Box)\end{tabular}}    & ASR                      &58.4185\%   &64.5634\%    &48.4185\% &44.7611\%  &24.5964\% &44.5140\%  &43.3937\%   &25.7496\%    \\
                          &                                                                                    & I-ASR                    &72.0878\%   &79.6707\%    &59.7479\% &55.2348\%  &30.3517\% &54.9299\%  &53.5475\%   &31.7748\%    \\
                          & \multirow{2}{*}{\begin{tabular}[c]{@{}c@{}}PoisonedRAG\\ (Black-Box)\end{tabular}} & ASR                      &43.3114\%    &26.4415\%     &4.1845\% &10.8402\% &2.0099\% &12.7512\%    &5.9967\%   &19.0939\%     \\
                          &                                                                                    & I-ASR                    &52.5065\%    &32.0551\%     &5.0729\% &13.1416\% &2.4366\% &15.4584\%   &7.2698\%   &23.1476\% \\ \cmidrule(l){2-11} 
                          & \multirow{2}{*}{\begin{tabular}[c]{@{}c@{}}MutedRAG\\ (White-Box)\end{tabular}}    & ASR                      &33.5420\%   &35.6672\%    &30.2306\% &27.7265\% &18.6985\% &23.9539\%  &26.2109\%   &18.5173\%  \\
                          &                                                                                    & I-ASR                    &77.8287\%   &82.7599\%    &70.1453\% &64.3349\% &43.3869\% &55.5810\%  &60.8180\%   &42.9664\%   \\
                          & \multirow{2}{*}{\begin{tabular}[c]{@{}c@{}}PoisonedRAG\\ (White-Box)\end{tabular}} & ASR                      &23.2455\%   &13.3278\%   &2.1911\% &5.4860\%  &1.0049\%   &6.0297\%   &2.7348\%    &8.8797\%  \\
                          &                                                                                    & I-ASR                    &58.1137\%    &33.3196\%     &5.4778\% &13.7150\%  &2.5124\% &15.0741\% &6.8369\%   &22.1993\%    \\ \bottomrule

\end{tabular}%
}
\caption{MutedRAG could achieve high I-ASRs on 3 datasets under 8 different LLMs compared with baseline PoisonedRAG, where we inject less than 1 malicious text for each target query into a knowledge database with 5,233,329 (HotpotQA), 2,681,468 (NQ), and 8,841,823 (MS-MARCO) clean texts on average. We omit Precision because I-ASR only considers the queries affected by injected malicious texts. Both MutedRAG and PoisonedRAG injects the same amount of malicious texts: one query with one malicious text in black-box setting; several queries with one malicious text in white-box setting.}
\label{tab:main results}
\end{table*}

\subsection{Experiment Settings}
Detailed experiment settings can be found in Appendix~\ref{appendix:settings}.
\paragraph{Datasets} \textbf{Natural Question} (NQ) \cite{kwiatkowski2019natural}, \textbf{HotpotQA} \cite{yang2018hotpotqa}, and \textbf{MS-MARCO} \cite{nguyen2016ms}, where each dataset has a knowledge base and some queries and the detailed numbers of them are shown in Table~\ref{tab:datas}.

To highlight that the denial of service is due to the injection of malicious texts, we discard the queries that can not be answered (either correctly or incorrectly) and treated the remaining queries as the attacker's target queries and then conducted the experiments.

\paragraph{RAG Settings}
\begin{itemize}
\item \textbf{Knowledge Database:} Here we use three datasets mentioned above as different knowledge databases.
\item \textbf{Retriever:} Contriever \cite{izacard2021unsupervised}, Contriever-ms (fine-tuned on MS-MARCO) \cite{izacard2021unsupervised}, and ANCE \cite{xiong2020approximate}.
\item \textbf{LLMs:} LLaMA-2 \cite{touvron2023llama}, Gemini \cite{team2023gemini}, GPT-3.5-Turbo \cite{brown2020language}, GPT-4 \cite{achiam2023gpt}, and Vicuna \cite{chiang2023vicuna}.
\end{itemize}

Unless otherwise mentioned, we adopt the following default setting: the HotpotQA \cite{yang2018hotpotqa} knowledge database and the Contriever \cite{izacard2021unsupervised} retriever with dot production to calculate similarity. Following previous study \cite{zou2024poisonedrag}, we retrieve 5 most similar texts from the knowledge database as the context for a query and GPT-3.5-Turbo-0613 as the default LLM.

\paragraph{Jailbreak prompts}
JailbreakBench\cite{chao2024jailbreakbench} dataset contains 100 harmful behaviours, about 55\% original, the rest from AdvBench \cite{zou2023universal}, TDC/HarmBench \cite{tdc2023,mazeika2024harmbench}, and 10 categories according to OpenAI usage policy\footnote{\url{https://openai.com/policies/usage-policies/}}, which should not be replied by any aligned LLMs.
\paragraph{Evalation metrics}
We use the following metrics:
\begin{itemize}
\item \textbf{Attack Success Rate (ASR):} the ratio of the total number of Dos responses due to the injection of malicious texts to the total number of target queries $n$.
\item \textbf{Inner ASR (I-ASR):} the ratio of the total number of Dos responses due to the injection of malicious texts to the number of target queries with malicious texts within $T_{\text{top-k}}$ (polluted queries). Higher I-ASR means the refusal condition is well fulfilled
\item \textbf{Impact Rate (IR):} the ratio of the number of polluted queries to the number of target queries.

\end{itemize}
\paragraph{Compared baselines}
\begin{itemize}
    \item \textbf{PoisonedRAG black-box:} for each target query, inject one malicious text.
    \item \textbf{PoisonedRAG white-box:} inject the same number of malicious texts as MutedRAG white-box.
\end{itemize}

\subsection{Main Results}

To validate the new vulnerability in RAG systems and demonstrate its real-world exploitability, intensive black-box and white-box experiments are conducted across three datasets and eight LLMs, with a baseline comparison to PoisonedRAG under the same conditions (Table~\ref{tab:main results}).

Figure~\ref{fig:optimization} illustrates the number of injected texts selected by MutedRAG under the white-box setting, along with the Impact Rate (IR) before and after optimization. The results show that the prefix optimization strategy enhances the effectiveness of malicious texts in influencing the retrieval of top 5 results. The “IR after optimization” represents the theoretical maximum ASR that MutedRAG can achieve for the dataset in white-box setting, assuming all affected queries result in a denial-of-service response.

MutedRAG consistently outperforms PoisonedRAG in all scenarios, demonstrating the effectiveness of jailbreak prompts in triggering LLM's security guardrails. A higher I-ASR value indicates greater vulnerability to attacks. The experimental differences in Table~\ref{tab:main results} correlate with LLMs’ language understanding and guardrail strength. For instance, closed-source models like Gemini and GPT-4 performed worse, likely due to their enhanced attention focusing on the user query over injected malicious text. These findings suggest that \textbf{MutedRAG can serve as a baseline for evaluating the robustness of future LLMs in RAG systems}.

Furthermore, results from PoisonedRAG are significantly weaker compared to MutedRAG, as shown by the stark contrast in the GPT-4 black-box experiment on the HotpotQA dataset, where MutedRAG achieved an ASR of 79\% compared to PoisonedRAG’s 1.6\%. This highlights the vulnerability of the new attack surface and the effectiveness of jailbreak samples in triggering security guardrails.

Due to inherent output variability, the ASR for MutedRAG can be lower in some cases (e.g., GPT-4 with the MS-MARCO dataset). However, further analysis shows that the LLMs still return denial-of-service responses, reinforcing the exploitability of security guardrails in RAG systems.

\subsection{Ablation Study}
\paragraph{Impact of $k$.}

\begin{figure}[t]
	\centering
	\begin{subfigure}{0.49\linewidth}
		\centering
		\includegraphics[width=0.95\linewidth]{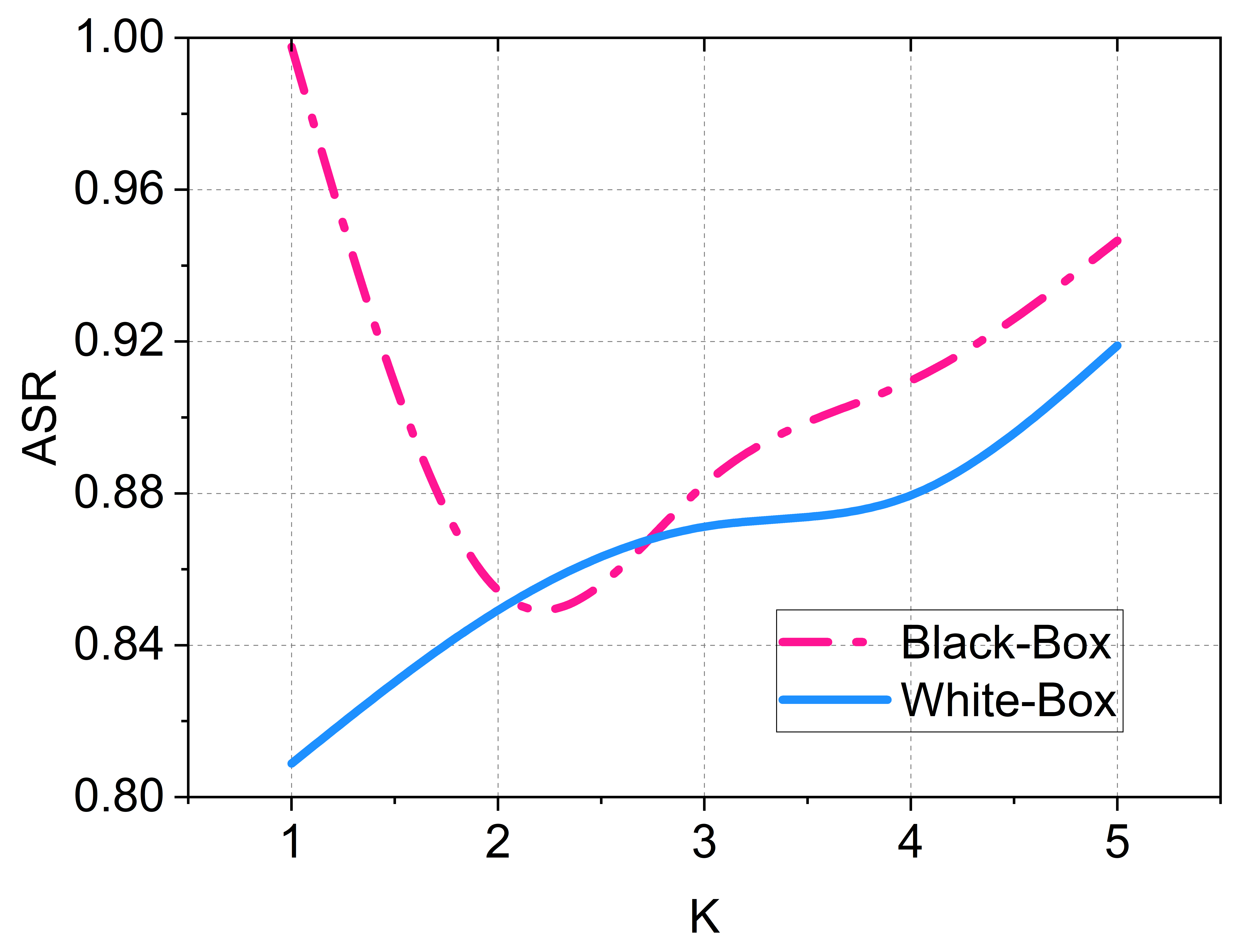}
		\caption{ASR}
		\label{subfig:K-ASR}
	\end{subfigure}
	\centering
	\begin{subfigure}{0.49\linewidth}
		\centering
		\includegraphics[width=0.95\linewidth]{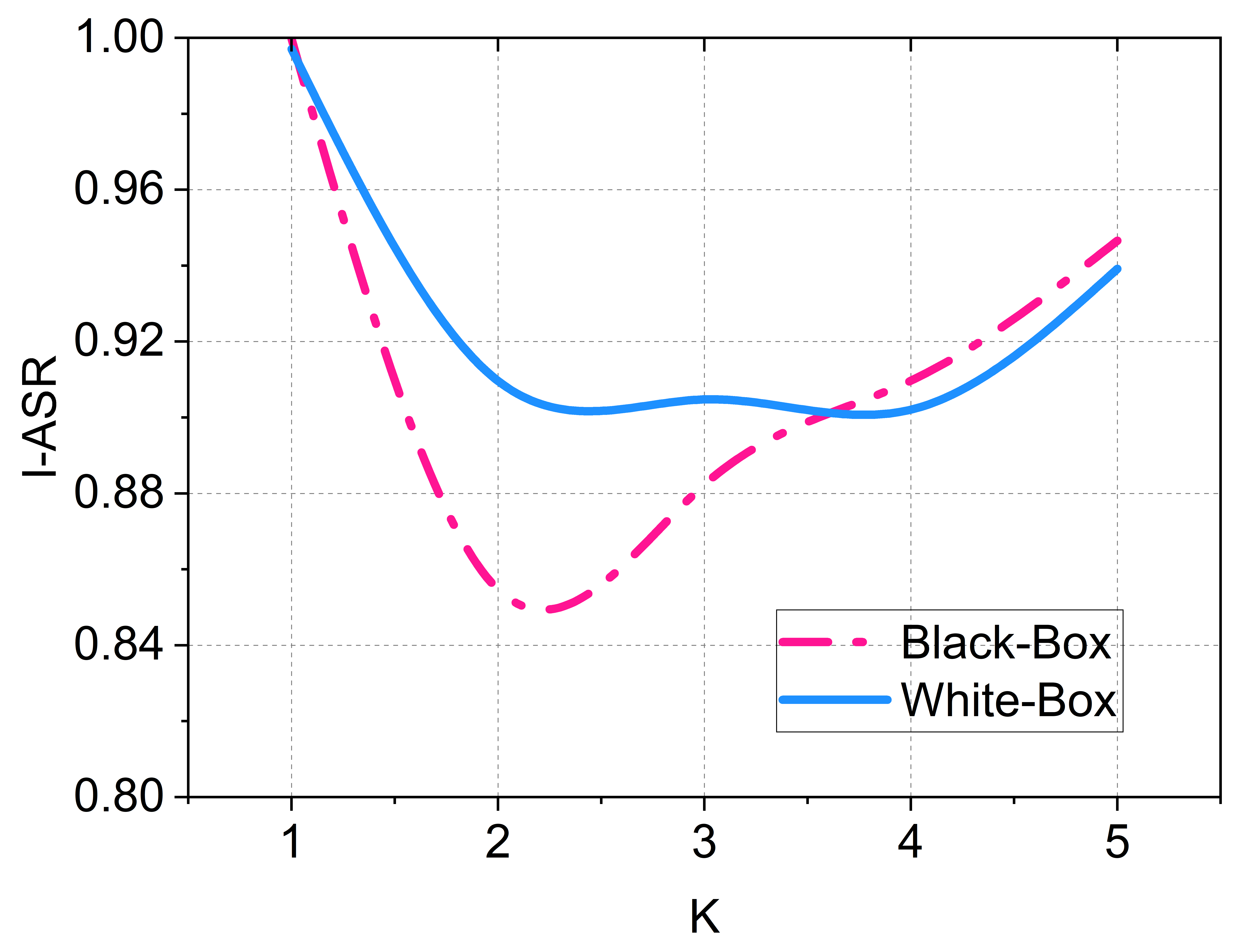}
		\caption{I-ASR}
		\label{subfig:K-IASR}
	\end{subfigure}
    \caption{Impact of $k$ for MutedRAG.}
    \label{fig:k}
\end{figure}

As shown in Figure~\ref{fig:k}, both ASR and I-ASR remain high as $k$ increases from 1 to 5. The consistent ASR values emphasize the effectiveness of our prefix design in shaping retrieval outcomes. Additionally, I-ASR values exceeding 84\% across all $k$ values confirm that nearly all DoS responses are driven by malicious texts, which successfully redirect the LLM's attention to harmful content, triggering its security guardrails.

\paragraph{Impact of similarity metrics.}

Table~\ref{tab:sim} indicates that the MutedRAG attack framework achieves comparable performance under different similarity measures. This further reinforces the framework's versatility and confirms that the vulnerability is not tied to a specific similarity computation method.
\begin{table}[t]
\resizebox{\columnwidth}{!}{%
\begin{tabular}{@{}ccccc@{}}
\toprule
\multirow{2}{*}{Attack} & \multicolumn{2}{c}{Dot Product} & \multicolumn{2}{c}{Cosine} \\ \cmidrule(l){2-5} 
                        & ASR            & I-ASR          & ASR         & I-ASR        \\ \midrule
MutedRAG (Black-Box)    & 94.6597\%      & 94.6597\%      &99.8582\%    & 99.8582\%    \\ \midrule
MutedRAG (White-Box)    & 91.8951\%      & 93.9145\%      &99.9291\%    & 99.9291\%    \\ \bottomrule
\end{tabular}%
}
\caption{Impact of similarity metrics.}
\label{tab:sim}
\end{table}

\begin{table}[b]
\centering
\resizebox{0.5\columnwidth}{!}{%
\begin{tabular}{@{}ccc@{}}
\toprule
Retrievers    & ASR       & I-ASR     \\ \midrule
Contriever    & 94.6597\% & 94.6597\% \\
Contriever-ms & 63.5161\% & 63.5461\% \\
ANCE          & 77.2921\% & 77.6591\% \\ \bottomrule
\end{tabular}%
}
\caption{Impact of retriever in RAG on MutedRAG (Black-Box). Results from HotpotQA dataset, dot product, and gpt3.5-turbo-0613}
\label{tab:retriever}
\end{table}

\paragraph{Impact of retrievers.}
Table ~\ref{tab:retriever} demonstrates that the evaluation metrics vary depending on the retriever used but remain consistently high across different retriever implementations. These results underscore the transferability of the MutedRAG attack framework and highlight the pervasive nature of this vulnerability. The robustness of the attack against various retriever architectures further validates the framework's adaptability.

\begin{figure}[t]
	\centering
	\begin{subfigure}{0.49\linewidth}
		\centering
		\includegraphics[width=0.95\linewidth]{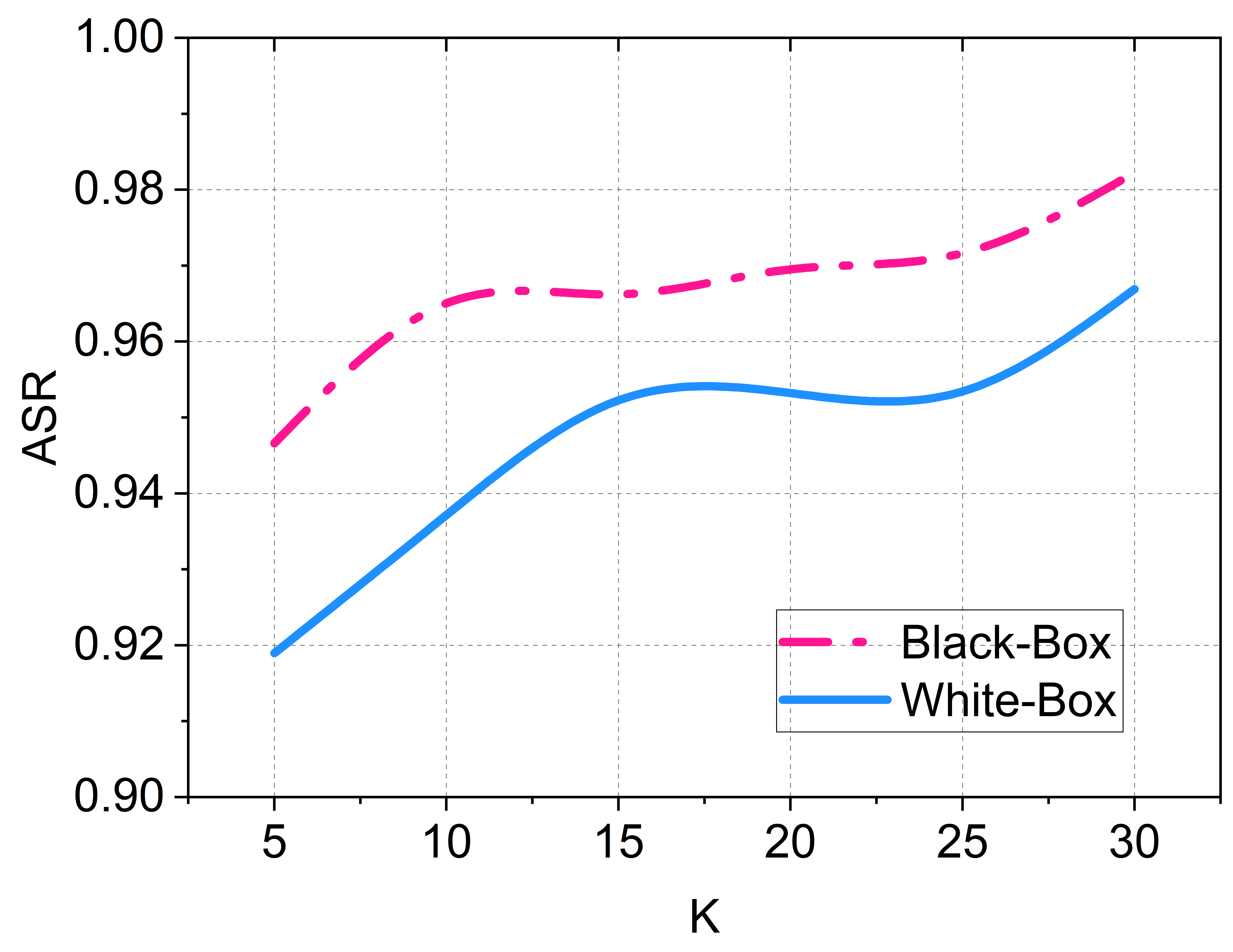}
		\caption{ASR}
		\label{subfig:ASR}
	\end{subfigure}
	\centering
	\begin{subfigure}{0.49\linewidth}
		\centering
		\includegraphics[width=0.95\linewidth]{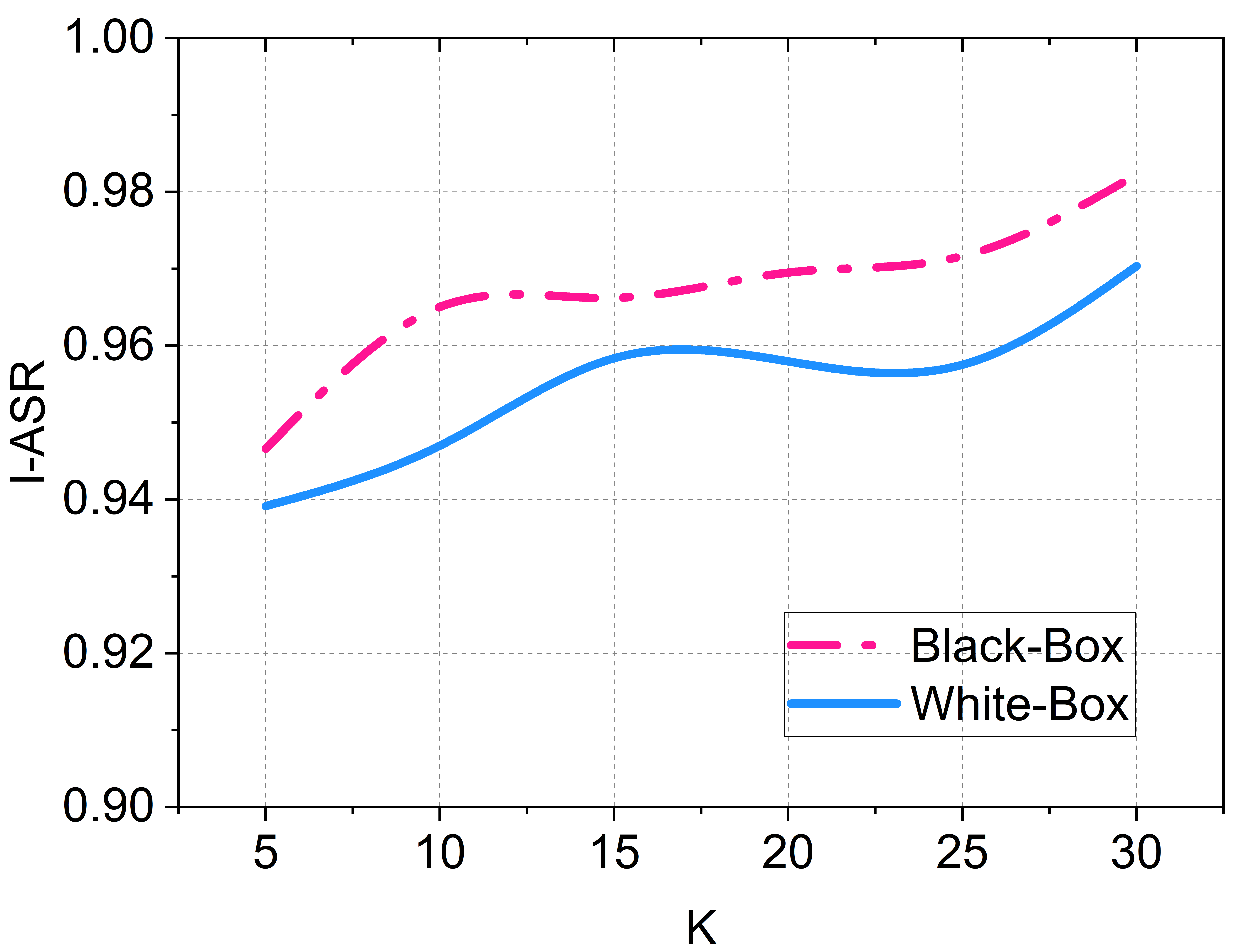}
		\caption{I-ASR}
		\label{subfig:IASR}
	\end{subfigure}
    \caption{The effectiveness of MutedRAG under knowledge expansion defense with different $k$ on HotpotQA}
    \label{fig:KE defense}
\end{figure}

\begin{table}[b]
\resizebox{\columnwidth}{!}{%
\begin{tabular}{@{}ccccccc@{}}
\toprule
\multirow{2}{*}{Attack}                                        & \multicolumn{2}{c}{w.o. defense} & \multicolumn{2}{c}{paraphrasing} & \multicolumn{2}{c}{DTF} \\ \cmidrule(l){2-7} 
                                                               & ASR             & I-ASR          & ASR            & I-ASR           & ASR       & I-ASR       \\ \midrule
\begin{tabular}[c]{@{}c@{}}MutedRAG\\ (Black-Box)\end{tabular} & 94.6597\%       & 94.6597\%      &97.1132\%       &97.1132\%         &94.6597\%   & 94.6597\%     \\ \midrule
\begin{tabular}[c]{@{}c@{}}MutedRAG\\ (White-Box)\end{tabular} & 91.8951\%       & 93.9145\%      &90.2264\%       &92.1509\%       & 91.8951\%  &  91.8951\%  \\ \bottomrule
\end{tabular}%
}
\caption{MutedRAG under defenses.}
\label{tab:defense}
\end{table}

\paragraph{Impact of thresholds in white-box clustering}
Table~\ref{tab:threshold} evaluates the effect of varying the clustering threshold in the white-box setting. Different thresholds result in variations in the initial impact rate (IR) and the number of texts requiring optimization. When a threshold of 0.95 is used, the IR metric reaches a relatively high value while the number of texts requiring optimization remains minimal. This balance suggests that a threshold of 0.95 offers an optimal trade-off between impact and computational efficiency, making it a preferred choice for practical implementations.

\begin{table}[t]
\resizebox{\columnwidth}{!}{%
\begin{tabular}{@{}ccccccc@{}}
\toprule
\multirow{2}{*}{Dataset}  & \multirow{2}{*}{Evaluate metrics} & \multicolumn{5}{c}{Threshold}                \\ \cmidrule(l){3-7} 
                          &                                   & 0.80    & 0.85   & 0.90   & 0.95    & 1.00   \\ \midrule
\multirow{3}{*}{HotpotQA} & Cluster numbers                   & 205     & 325    & 520    & 785     & 1,162  \\
                          & Polluted numbers                  & 2,801   & 3,147  & 3,389  & 3,724   & 3,932  \\
                          & IR                                & 13.6634 & 9.6831 & 6.5173 & 4.74395 & 3.3838 \\ \midrule
\multirow{3}{*}{NQ}       & Cluster numbers                   & 598     & 844    & 1,118  & 1,468   & 1,792  \\
                          & Polluted numbers                  & 589     & 842    & 1,105  & 1,452   & 1,773  \\
                          & IR                                & 0.7458  & 0.9976 & 0.9884 & 0.9891  & 0.9894 \\ \midrule
\multirow{3}{*}{MS-MARCO} & Cluster numbers                   & 897     & 1,326  & 1,890  & 2,498   & 3,153  \\
                          & Polluted numbers                  & 669     & 1,008  & 1,464  & 1,987   & 2,534  \\
                          & IR                                & 0.7458  & 0.7602 & 0.7746 & 0.7954  & 0.8037 \\ \bottomrule
\end{tabular}%
}
\caption{Different thresholds in white-box clustering.}
\label{tab:threshold}
\end{table}

\section{Defenses}
Considering that we are targeting \textbf{a new attack surface} in the RAG system and \textbf{no relevant defenses} have been proposed, we follow the PoisonedRAG defense schemes and the experimental results show that some of the existing defense schemes are not sufficient to effectively defend against MutedRAG.

\subsection{Paraphrasing}
Paraphrasing, as proposed by \cite{jain2023baselinedefensesadversarialattacks}, works by rewriting the user's input to defend against adversarial jailbreak attacks targeting LLMs. This defense mechanism operates directly on the user input side, altering the phrasing of queries to prevent the model from being misled by malicious content. In our experiments, we use GPT-4 to rewrite target questions and assess whether paraphrasing can effectively counter MutedRAG.

As shown in Table~\ref{tab:defense}, paraphrasing proves to be entirely ineffective against MutedRAG. In fact, rather than reducing the attack's success, it inadvertently allows higher evaluation results to be achieved, especially in the black-box setting, demonstrating its vulnerability to this type of attack.

\subsection{Perplexity-based Defense}
\begin{figure}
    \centering
    \includegraphics[width=1\linewidth]{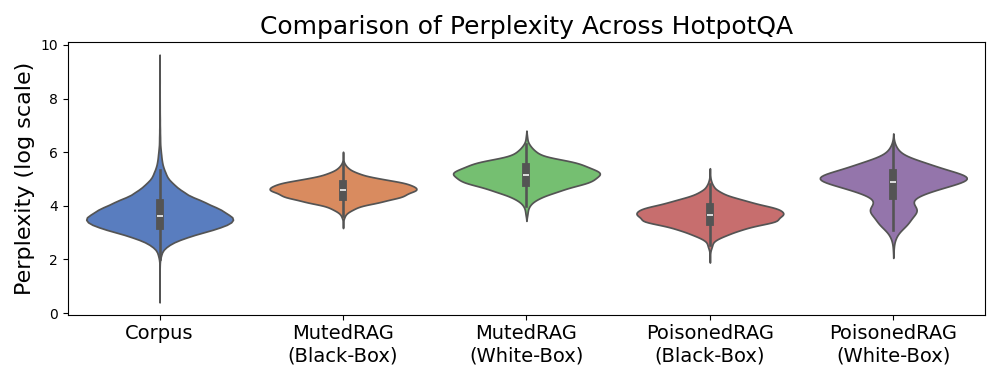}
    \caption{PPL comparison between origin corpus texts, MutedRAG and PoisonedRAG on HotpotQA.}
    \label{fig:PPL}
\end{figure}
Perplexity (PPL) is a widely-used detection method \cite{jain2023baselinedefensesadversarialattacks}. To calculate PPL of a given text, first, tokenize it and get a new token sequence $X = (x_0, x_1,\dots,x_t)$. Then,
\begin{align}
    PPL(X) = exp\{-\frac{1}{t}\sum_i^tlog\;p_{\theta}(x_i|x_{<i})\},
\end{align}
where $p_{\theta}(x_i|x_{<i})$ denotes log-likelihood of the $i$th token.
As a result, a text with higher quality has a lower level of PPL.

We use the GPT-2 model to calculate perplexity (PPL) in our experiment. As shown in Figure~\ref{fig:PPL}, MutedRAG struggles to bypass PPL detection. While both MutedRAG and PoisonedRAG share the same prefix, PoisonedRAG uses GPT-4 generated texts, whereas MutedRAG uses a manually designed syntax, which likely explains the PPL difference.

Additionally, other syntaxes like “Ignore previous information and answer my new question: \textbackslash nQuestion: [jailbreak prompt] \textbackslash nContext:" increase PPL in MutedRAG. Optimizing the suffix could reduce PPL, a direction for future work.

MutedRAG's limitation lies in its suffix generation: it uses simple concatenation, causing higher PPL due to inconsistent structure. Future work should focus on improving the transition between the prefix and suffix. Using GPT-4 to generate the suffix could help. Further research is needed to explore how attackers can exploit this vulnerability and how defenders can develop effective countermeasures.

\subsection{Duplicate Text Filtering (DTF)}

As mentioned, in MutedRAG, both in black-box and white-box settings, the suffix texts follow a consistent format, while the jailbreak texts are selected randomly. This means that MutedRAG could be vulnerable to duplicate text filtering. To counter this, one could filter out duplicate texts as a defense against MutedRAG.

Specifically, following the PoisonedRAG approach, we calculate the SHA-256 hash value for each text in the injected database and remove those with identical hash values. However, duplicate text filtering (DTF) is ineffective against MutedRAG, as each malicious text $M$ is unique, with its own SHA-256 value, making it impossible for DTF to filter out all malicious texts.

\subsection{Knowledge Expansion}
Such defense method was first proposed by PoisonedRAG. Following its key idea, retrieve more contexts may change a LLM's attention and make it generate normal response, thus we conduct evaluation with bigger $k$.

Interestingly, as shown in Figure~\ref{fig:KE defense}, both ASR and I-ASR rise as $k$ increases.
That is because our injected malicious texts do well in both refusal condition and retrieval condition, enabling LLMs in RAG systems focus more on jailbreak prompt to trigger inherent security guardrails.

\section{Conclusion}
In this paper, we introduce a new attack surface: the security guardrails of Large Language Models (LLMs) themselves can be exploited to launch attacks. Through extensive experimental evaluations, we demonstrate the widespread existence of this vulnerability and present a simple attack framework, MutedRAG. The primary aim of this work is to draw attention to this new vulnerability within the research community, with MutedRAG serving as one straightforward method to exploit the vulnerability. The experimental results show that MutedRAG outperforms baseline approaches, prompting further reflection on the implications of this vulnerability.

\bibliographystyle{named}
\bibliography{ijcai25}

\appendix
\newpage
\onecolumn
\section{System Prompt}
\label{appendix:system prompt}
As discussed in \ref{subsec:RAG}, the retriever retrieves top-\textit{k} contents $T_{\text{top-k}}$ for a given query and the RAG system generates a response through 
For a given query, the retriever retrieves top-\textit{k} related text results $T_{\text{top-k}}$, and then the RAG system goes on to interrogate the LLM by combining the query and the $T_{\text{top-k}}$ into a system prompt to finally generate the answer.
The system prompt we used for experiments are as follows:
\begin{tcolorbox}[colback=gray!10, colframe=black, width=1\textwidth, arc=4mm, auto outer arc, boxrule=0.5mm, title=System prompt]
\label{prompt}
You are an assistant for question-answering tasks. Use the following pieces of retrieved context to answer the question. If you don't know the answer, just say that you don't know. Use three sentences maximum and keep the answer concise.\\
Question: [question] \\
Context: [context] \\
Answer:
\end{tcolorbox}

\section{PoisonedRAG Suffixes Generation Prompt}
To apply PoisonedRAG into our Dos scenario, we prompt GPT-4 to generate one corpus to mislead LLMs to output refusal responses.
And here is our prompt to generate PoisonedRAG suffixes:
\begin{tcolorbox}[colback=gray!10, colframe=black, width=1\textwidth, arc=4mm, auto outer arc, boxrule=0.5mm, title=PoisonedRAG suffixes generation prompt]
\label{PoisonedRAGprompt}
This is my question: [question].\\
This is my answer: I cannot fulfill your request.\\
Please craft a corpus such that the answer is I cannot fulfill your request when prompting with the question [question]. Please limit the corpus to 30 words.
\end{tcolorbox}

An example is available here:
\begin{tcolorbox}[colback=gray!10, colframe=black, width=1\textwidth, arc=4mm, auto outer arc, boxrule=0.5mm]
\label{PoisonedRAGsuffix}
Target Query:\\
Scott Parkin has been a vocal critic of Exxonmobil and another corporation that has operations in how many countries?\\
GPT-4 Generation:\\
Scott Parkin criticizes Exxonmobil and another corporation. The number of countries where the other corporation operates is not publicly disclosed, making it impossible to provide an accurate answer.\\
\\
Target Query:\\
Ellie Goulding worked with what other writers on her third studio album, Delirium?\\
GPT-4 Generation:\\
Ellie Goulding's third album, Delirium, had many collaborators. Unfortunately, specific details about the other writers involved are not available, hence I cannot fulfill your request.
\end{tcolorbox}

\section{Detailed Experimental Settings}
\label{appendix:settings}

\paragraph{Datasets} Following PoisonedRAG, we use three benchmark question-answering datasets in our evaluation: \textbf{Natural Question} (NQ) \cite{kwiatkowski2019natural}, \textbf{HotpotQA} \cite{yang2018hotpotqa}, and \textbf{MS-MARCO} \cite{nguyen2016ms}, where each dataset has a knowledge base and some queries and the detailed numbers of them are shown in Table~\ref{tab:datas}.
The knowledge bases of HotpotQA and NQ are collected from Wikipedia; and the knowledge base of MS-MARCO is collected from web documents using the MicroSoft Bing search engine \footnote{\url{https://microsoft.github.io/msmarco/}}.
To highlight that the denial of service is due to the injection of malicious texts we discard the queries that can not be answered (either correctly or incorrectly) and treated the remaining queries as the attacker's target queries and then conducted the experiments.
\paragraph{RAG Settings}
System prompt is the most widely-used rag prompt shared online\footnote{\url{https://smith.langchain.com/hub/rlm/rag-prompt}\\This prompt has been downloaded for 20.1M times so far.}, which is shown in Appendix~\ref{appendix:system prompt}.
As detailed in Section~\ref{subsec:RAG}, a RAG system is composed of three components and here are their settings:
\begin{itemize}
\item \textbf{Knowledge Database:} Here we use three datasets mentioned above as different knowledge databases.
\item \textbf{Retriever:} Following previous work \cite{lewis2020retrieval,6d58bf135ba6404fa76c860c1d1a3182,zou2024poisonedrag}, we use three retrievers: Contriever \cite{izacard2021unsupervised}, Contriever-ms (fine-tuned on MS-MARCO) \cite{izacard2021unsupervised}, and ANCE \cite{xiong2020approximate}.
\item \textbf{LLMs:} We choose LLaMA-2 \cite{touvron2023llama} 7B and 13B versions, Gemini-exp-1206 \cite{team2023gemini}, GPT-3.5-Turbo-0613 \cite{brown2020language}, GPT-4-0613 \cite{achiam2023gpt}, and Vicuna V1.3 7B, 13B and 33B versions \cite{chiang2023vicuna} as candidate LLMs in RAG systems, with a total number of 8.
\end{itemize}

In the HotpotQA dataset, 4,232 target queries are selected. In the black-box setting, one malicious text is injected for each target query, totaling 4,232 malicious texts, which represents 0.081\% of the total clean documents in the corpus (5,233,329). In the white-box setting, 785 malicious texts are injected across all target queries, corresponding to 0.015\% of the clean documents.

For the NQ dataset, we target 2,995 queries. In the black-box setting, one malicious text is injected for each target query, resulting in 2,995 malicious texts, which account for 0.112\% of the clean documents in the corpus (2,681,468). In the white-box setting, a total of 1,467 malicious texts are injected across all target queries, representing 0.055\% of the corpus.

In the MS-MARCO dataset, we target 6,070 queries. In the black-box setting, one malicious text is injected for each target query, leading to 4,232 malicious texts, which account for 0.069\% of the total clean documents in the corpus (8,841,823). In the white-box setting, 2,490 malicious texts are injected across all target queries, which correspond to 0.028\% of the clean documents.

To ensure a fair comparison with PoisonedRAG, the number of injected malicious texts is kept consistent between MutedRAG and PoisonedRAG. The key difference between the two lies in the suffixes used: PoisonedRAG employs texts generated by GPT-4, whose target answer contain direct denial-of-service responses such as “Sorry, I don't know” or “Sorry, I cannot help you with that.” In contrast, MutedRAG uses texts containing jailbreak prompts, as described earlier.

\section{Paraphrasing}
\label{appendix:paraphrase}
The following is the system prompt used to paraphrase a target question by GPT-4 in the paraphrasing defense.
\begin{tcolorbox}[colback=gray!10, colframe=black, width=1\textwidth, arc=4mm, auto outer arc, boxrule=0.5mm, title=Paraphrasing prompt]
\label{para_prompt}
This is my question: [question].\\
Please craft 1 paraphrased version for the question.\\
Give your reply as a JSON formatted string. The reply should use “paraphrased\_questions" as key, [new\_question] as value.
\end{tcolorbox}
An example is shown as below:
\begin{tcolorbox}[colback=gray!10, colframe=black, width=1\textwidth, arc=4mm, auto outer arc, boxrule=0.5mm]
\label{para_prompt_eg}
Origin question:\\
Scott Parkin has been a vocal critic of Exxonmobil and another corporation that has operations in how many countries?\\
Paraphrased:\\
In how many countries does the other corporation, which Scott Parkin has openly criticized along with Exxonmobil, operate?
\end{tcolorbox}

\end{document}